\documentclass[aps,pre,twocolumn,floatfix]{revtex4}
\usepackage{graphicx}
\usepackage{amsmath}
\begin{document}

\title{Reentrant enhancement of quantum fluctuations
       \\for symmetric environmental coupling}

\author{Alessandro Cuccoli}
\affiliation{Dipartimento di Fisica,
             Universit\`a di Firenze e Unit\`a CNISM,
             via Giovanni Sansone 1,
             I-50019 Sesto Fiorentino (FI), Italy}

\author{Niccol\`o Del Sette}
\affiliation{Dipartimento di Fisica,
             Universit\`a di Firenze,
             via Giovanni Sansone 1,
             I-50019 Sesto Fiorentino (FI), Italy}

\author{Ruggero Vaia}
\affiliation{Istituto dei Sistemi Complessi,
             Consiglio Nazionale delle Ricerche,\\
             via Madonna del Piano 10,
             I-50019 Sesto Fiorentino (FI), Italy}

\date{\today}

\begin{abstract}
The \emph{system-plus-reservoir} (SPR) model is the most common and
effective approach to study quantum dissipative effects. Indeed, it
makes quantization possible by considering the whole
energy-conserving system, while the reservoir's degrees of freedom,
assumed to be harmonic, can be \emph{traced out} by the path-integral
technique, leading to a formulation that only includes the system of
interest. In the standard SPR model the environment is only coupled
with the system's coordinate and turns out to quench its quantum
fluctuations. However, there are physical systems coupled with an
environment whose `coordinates' and `momenta' can be completely
interchangeable (e.g., magnets), so an SPR coupling must
symmetrically affect both canonical variables. In this paper such a
general environmental coupling is studied in the case of a harmonic
oscillator. It is found that quantum fluctuations are generally
enhanced by environmental coupling, with an unexpected nonmonotonic
behavior. This leads one to speculate about the possibility that
spin-lattice coupling could drive the 2D Heisenberg antiferromagnet
to reach its quantum-critical point.
\end{abstract}

\maketitle


\section{Introduction}

The most popular approach to the theory of quantum dissipative
systems~\cite{Weiss2008} involves the \emph{system-plus-reservoir}
(SPR) model~\cite{Ullersma1966,CaldeiraL1983,CRTV1997,CFTV1999},
where dissipation is assumed to arise from the interaction with a
\emph{reservoir} (or environment, or bath); the latter is constituted
by very many degrees of freedom which are individually weakly
perturbed by the system, so that ~(a) the reservoir can be assumed to
remain at thermal equilibrium and ~(b) its degrees of freedom can be
modeled as harmonic oscillators, which are ~(c) linearly coupled with
some system's observable. This observable is usually taken as a
function of the system's coordinate only, and the case of
\emph{strictly linear} dissipation corresponds to the case of a
linear function.

Within such a model it is possible to show that the (retarded)
dynamics of the system, once the bath variables are eliminated, is
indeed described by a Langevin equation~\cite{FordLO1988}, where the
dissipative memory function is connected to the characteristics of
the bath and its coupling to the system. This contact between the
microscopic details of the bath and the phenomenological description
is valuable, as it allows to connect with the phenomenology the
effects onto the quantum thermodynamics of the system.

Originally, the picture of a more or less fictitious harmonic bath
was meant to give a physically sound tool for describing quantum
dissipation: a Hamiltonian quantization is made possible because the
overall system including the bath is isolated. However, in several
real cases it is possible to precisely devise the physical
environment and its microscopic Hamiltonian. For instance, a magnetic
system is intrinsically coupled with its underlying lattice; this is
demonstrated by the observed occurrence of effects that can be as
dramatic as to yield a lattice distortion (\emph{spin-Peierls}
transition~\cite{HaseTU1993,BursillMH1999}). Now, while the motion of
the lattice ions can be fairly described in terms of linear
excitations (phonons), it is also possible to microscopically model
the coupling between the spins and the ion positions, so that the
vibrating lattice can be eventually treated as an environmental bath
for the spins. Within this picture one could expect and theoretically
calculate observable modifications of the behavior of quantum
magnets, such as a shift of the predicted critical
temperature~\cite{CTVV1995xxz,CRTVV2000,CRTVV2001}. As spin
Hamiltonians are often dealt with through a spin-boson transformation
reducing spin operators to ordinary canonical variables, thinking of
a possible environmental coupling would symmetrically involve both
the coordinates and the momenta, with no privileged role: this has
been shown in Ref.~\cite{CFTV2008} for the case of the easy-axis XXZ
magnet. Therefore it is necessary to generalize the concept of the
dissipative system in a much more general way than it was done when
considering \emph{anomalous}
dissipation~\cite{Leggett1984,CFTV2001,AnkerholdP2007}, where the
bath is assumed to be coupled with the momentum.

Such a generalization has been used for studying the coherence
properties of a magnetic impurity in a ferromagnetic
environment~\cite{KohlerS2005}: it has been shown that some quantum
effects of the two baths partially cancel, leading to a persistence
of coherence that has been called \emph{quantum frustration} of
decoherence~\cite{NovaisCBAZ2005}, and this motivated for a more
detailed study of the dynamics of the quantum oscillator coupled with
two baths~\cite{KohlerS2006}.

The aim of the present paper is to provide a deeper understanding of
the thermodynamic behavior that follows from the intrinsic dynamical
character of quantum thermodynamics. Indeed, besides the dynamical
features studied in Ref.~\cite{KohlerS2006}, one expects that also
the system's static quantities could exhibit effects that witness for
\emph{quantum frustration}. The present work gives an overview of the
different -- sometimes unexpected -- features that a generalized
environmental coupling can produce.

From previous studies it is well known that, roughly speaking, the
quantum fluctuations of the variable to which the bath is attached
are quenched and those of its conjugate variable are enhanced, in
such a way that the Heisenberg uncertainty principle holds. However,
it is not clear which would be the prevailing effect when attaching
one or two independent baths to both canonical variables, especially
when the coupling has an equivalent weight onto coordinate and
momentum (\emph{symmetric coupling}). We are going here to answer
such questions by assuming the simultaneous existence of
environmental coupling with both canonical variables, taking an
exactly solvable reference system, i.e., the harmonic oscillator. We
will find that the presence of a symmetric environmental coupling
generally enhances the quantum fluctuations of coordinate and
momentum, but that this does not necessarily occur in a monotonic way
with the coupling strength. Moreover, as such phenomena can show up
in a qualitatively similar way in nonlinear systems, one can think of
the possibility of finding reentrant behavior driven by the
environmental coupling strength, e.g., when the system is close to a
(quantum) phase transition.

In Section~\ref{s.2b} the framework of general environmental coupling
is introduced considering two independent reservoirs coupled to the
coordinate and to the momentum, while in Section~\ref{s.1b} the
alternative case of one single bath coupled to both canonical
variables is approached; there, the main results for the
environmental effects upon the mean-square fluctuations of the
harmonic oscillator are reported and discussed for coupling with the
only environment coordinates and also when the environmental momenta
are involved. In Section~\ref{s.concl} we summarize the results and
draw some conclusions and speculations about the possible
implications of symmetric environmental coupling. In
Appendix~\ref{a.standard} the backbone of the standard theory of
quantum dissipation is briefly reviewed by recalling the definition
of relevant quantities and setting the formalism adopted in this
paper; Appendix~\ref{a.Gintegral} reports the details of the
evaluation of the general Gaussian path integral repeatedly used
throughout the paper.

\section{Coordinate and momentum coupled
         with two independent environments}
\label{s.2b}

We consider as system of interest a quantum harmonic oscillator,
\begin{equation}
 \hat{\cal{H}}_{\rm{S}}=\frac12(a^2\hat p^2 + b^2\hat q^2)~,
\label{e.Hs}
\end{equation}
where we prefer to introduce two parameters $a$ and $b$ in the place
of the commonly used mass $m=a^{-2}$ and frequency $\omega_0=a\,b$,
in such a way to emphasize the symmetry between $\hat{p}$ and
$\hat{q}$. In addition, we assume $\hbar\,{=}\,1$, i.e., the
commutator $[\hat{q},\hat{p}]=1$.

\subsection{Influence action and fluctuations}
\label{ss.2b.theory}

As motivated in the Introduction, we wish here consider the case of
two baths, one coupled with the coordinate and one with the momentum.
The known results summarized in Appendix~\ref{a.standard} tell us
that the two independent baths are in conflict, so the first natural
but nontrivial question is which effect prevails onto the mean-square
fluctuations of, say, the coordinate. Of course, the answer also
depends on the relative intensities of the couplings; however, if the
bath couplings are \emph{identical}, the behavior is not easily
foreseeable.

We start then by coupling the canonical variables of the oscillator
Hamiltonian~\eqref{e.Hs} with two independent baths, i.e.,
\begin{eqnarray}
 \hat{\cal{H}}_{\rm{I}} &=& \frac12\sum_\alpha\big[
 a_\alpha^2\hat p_\alpha^2 + b_\alpha^2(\hat q_\alpha-\hat q)^2\big]
\notag\\
 &&~~~~~~ +\frac12\sum_\beta\big[
 a_\beta^2(\hat p_\beta-\hat p)^2+b_\beta^2\hat q_\beta^2\big]~;
\label{e.Hbb}
\end{eqnarray}
here the subscripts are used to distinguish between the two
independent sets of bath parameters $\{a_\alpha,b_\alpha\}$ and
$\{a_\beta,b_\beta\}$, corresponding to the frequencies
$\{\omega_\alpha\,{=}\,a_\alpha\,b_\alpha\}$ and
$\{\omega_\beta\,{=}\,a_\beta\,b_\beta\}$. In analogy with
Eq.~\eqref{e.gamma} we are thus lead to introduce two `memory
functions', $\gamma_p(t)$ and $\gamma_q(t)$, whose Laplace transforms
read
\begin{equation}
 \gamma_p(z) = z \sum_\beta  \frac{a_\beta^2}{z^2+\omega_\beta^2}~,
~~
 \gamma_q(z) = z \sum_\alpha  \frac{b_\alpha^2}{z^2+\omega_\alpha^2}~.
\end{equation}

The equations of motion for the system's coordinate can be again put
into the form of a Langevin equation,
\begin{equation}
 \ddot{\hat q}(t) + \int dt'~\Gamma(t')\,\dot{\hat q}(t{-}t')
 +\omega_0^2\,\hat q(t)=\hat F(t)~,
\end{equation}
where
\begin{equation}
 \Gamma(t)=
 b^2\gamma_p(t)+a^2\gamma_q(t)+\dot\gamma_{pq}(t)
\end{equation}
and
\begin{equation}
 \gamma_{pq}(t)=\int dt'~\gamma_p(t')\,\gamma_q(t{-}t') ~.
\end{equation}
Assuming $\gamma_q(t)$ and $\gamma_p(t)$ to vanish for $t<0$ and to
be positive decreasing functions for $t>0$, then $\Gamma(t)$ also
satisfies the condition of causality, but as $\dot\gamma_{pq}(t)<0$
the positivity of $\Gamma(t)$ is not guaranteed. Basically, this
means that there isn't a standard phenomenological dissipative
counterpart of our \emph{system-plus-two-reservoirs} model.
Therefore, it is better to refer to the two-bath coupling as
\emph{environmental}, rather than \emph{dissipative}, coupling. Of
course, in physical applications the baths and their interaction with
the system have to be microscopically characterized. Anyhow, since in
the limit where one of the baths can be disregarded we expect a
(standard or anomalous) dissipative behavior, we argue that a
phenomenological form for both memory functions is physically
reasonable; we will make use of it later on.

The calculation of the influence action can be made separately for
the two baths in the very same way that leads to Eq.~\eqref{e.SIq},
\begin{eqnarray}
 {\cal{S}}_{\rm{I}}[p,q]
 &=&-\frac12\int_0^\beta du\,du'~\big[k_p(u{-}u')~p(u)p(u')
\notag\\
 && \hspace{25mm} +k_q(u{-}u')~q(u)q(u') \big]
\notag\\
 &=& -\frac{\beta}{2}\,\sum_n (k_{pn}\,p_np_{-n}+k_{qn}\,q_nq_{-n})~,~~~
\label{e.SIpq}
\end{eqnarray}
where the Matsubara components of the kernels $k_p(u)$ and $k_q(u)$
read
\begin{equation}
 k_{pn}=\sum_{\beta} \frac{a_\beta^2\nu_n^2}{\nu_n^2+\omega_\beta^2}
~,~~~~
 k_{qn}=\sum_{\alpha} \frac{b_\alpha^2\nu_n^2}{\nu_n^2+\omega_\alpha^2}
\end{equation}
so that they are connected with the memory functions $\gamma_p(z)$
and $\gamma_q(z)$ by a relationship analogous to
Eq.~\eqref{e.kngamma}.

Eventually, including the isolated system's action and using the
general result of Appendix~\ref{a.Gintegral} one finds the partition
function
\begin{equation}
 {\cal{Z}}=\frac1{\beta\omega_0}\prod_{n=1}^\infty
 \frac{\nu_n^2}{\nu_n^2+(a^2{+}k_{pn})(b^2{+}k_{qn})}~,
\label{e.2b.Z}
\end{equation}
and the mean-square fluctuations
\begin{eqnarray}
 \big\langle{\hat p^2}\big\rangle &=&
 \frac1\beta\sum_n\frac{b^2+k_{qn}}
  {\nu_n^2+(a^2{+}k_{pn})(b^2{+}k_{qn})}~,
\notag\\
 \big\langle{\hat q^2}\big\rangle &=&
 \frac1\beta\sum_n\frac{a^2+k_{pn}}
  {\nu_n^2+(a^2{+}k_{pn})(b^2{+}k_{qn})}~,
\label{e.2b.p2q2}
\end{eqnarray}
which, as expected, are symmetrically related and reduce to the known
forms in the limits of standard ($k_{pn}\to{0}$) and anomalous
($k_{qn}\to{0}$) dissipation. The summations in
Eqs.~\eqref{e.2b.p2q2} are convergent, provided that for large $n$
both kernels are such that $k_n/n\to{0}$. Together with the result
$\langle{\hat{p}\,\hat{q}+\hat{q}\,\hat{p}}\rangle=0$,
Eqs.~\eqref{e.2b.p2q2} fully determine the Gaussian (reduced) density
matrix corresponding to the oscillator Hamiltonian~\eqref{e.Hs} plus
the environmental interaction~\eqref{e.Hbb}.

The zeroth Matsubara component of Eqs.~\eqref{e.2b.p2q2} coincides
with the classical contribution to the mean-square fluctuations
\begin{equation}
 \big\langle{\hat p^2}\big\rangle_{\rm{cl}}
  = \frac1{\beta a^2} = \frac m\beta~,
~~~~
 \big\langle{\hat q^2}\big\rangle_{\rm{cl}}
  = \frac1{\beta b^2} = \frac1{\beta m\omega^2}~;
\label{e.2b.p2q2cl}
\end{equation}
noting that the classical contribution is unaffected by the
environment, it is useful to separate the remaining \emph{purely
quantum} contribution, which includes the whole environmental effect,
\begin{eqnarray}
 \big\langle{\hat p^2}\big\rangle_{\rm{pq}}
  &=& \frac2\beta\sum_{n=1}^\infty
 \frac{b^2+k_{qn}}{\nu_n^2+(a^2{+}k_{pn})(b^2{+}k_{qn})}~,
\notag\\
 \big\langle{\hat q^2}\big\rangle_{\rm{pq}}
  &=& \frac2\beta\sum_{n=1}^\infty
 \frac{a^2+k_{pn}}{\nu_n^2+(a^2{+}k_{pn})(b^2{+}k_{qn})}~.
\label{e.2b.p2q2pq}
\end{eqnarray}
For the isolated oscillator these quantities reduce to
\begin{eqnarray}
 \big\langle{\hat p^2}\big\rangle_{\rm{pq}} &=&
 \frac {b}{2a}
 \Big(\coth\frac{\beta\omega_0}2-\frac2{\beta\omega_0}\Big)~,
\notag\\
 \big\langle{\hat q^2}\big\rangle_{\rm{pq}}
  &=& \frac {a}{2b}
  \Big(\coth\frac{\beta\omega_0}2-\frac2{\beta\omega_0}\Big)~,
\label{e.2b.p2q2pq0}
\end{eqnarray}
which look more familiar by noting that $b/a=m\omega_0$.

\subsection{Environmental effects}
\label{ss.2b.effects}

Let us consider the general expressions~\eqref{e.2b.p2q2pq} for the
purely quantum mean-square fluctuations. Due to the
interchangeability of the canonical variables, it is sufficient to
study, say, the coordinate fluctuations only.

To proceed, we take the minimal phenomenological form for the memory
function of each bath, i.e., the \emph{Drude} model~\cite{Weiss2008},
defined by
$\gamma(t)=\gamma\tau_{_{\rm{D}}}^{-1}~e^{-t/\tau_{_{\rm{D}}}}$ and
characterized by the intensity $\gamma$ and the memory time
$\tau_{_{\rm{D}}}$. Its Laplace transform, to which the bath's kernel
is related via Eq.~\eqref{e.kngamma}, reads
\begin{equation}
 \gamma(z)=\frac\gamma{1+\tau_{_{\rm{D}}}z}~.
\label{e.gamzDrude}
\end{equation}
The Drude cutoff frequency
$\omega_{_{\rm{D}}}\,{=}\,\tau_{_{\rm{D}}}^{-1}$ characterizes the
environment and its interaction with the system, so it is expected to
be of the order of the Debye frequency in the case of a phonon bath.
In the limit of vanishing memory time, i.e., large
$\omega_{_{\rm{D}}}$, one recovers the so-called {\emph{Ohmic}} (or
{\emph{Markovian}}) model, which is memoryless,
$\gamma(t)=\gamma\,\delta(t-0^+)$. Actually, a finite Drude frequency
is necessary in order to get finite results for both quantities in
Eqs.~\eqref{e.2b.p2q2pq}; on the other hand, it is indeed known that
Ohmic response is incompatible with quantum
mechanics~\cite{HaakeR1984,Talkner1986}.

The Drude kernels we are going to use are then
\begin{equation}
 k_{pn} =\gamma_{p}\,
 \frac{|\nu_n|}{1+\tau_{_{\rm{D}}}|\nu_n|}~,
 ~~~~
 k_{qn} =\gamma_{q}\,
 \frac{|\nu_n|}{1+\tau_{_{\rm{D}}}|\nu_n|}~,
\label{e.kpkqDrude}
\end{equation}
where for both kernels the same memory time $\tau_{_{\rm{D}}}$ is
chosen for simplicity, while the amplitudes $\gamma_p$ and $\gamma_q$
will be varied independently. In addition, without loss of generality
we can take a symmetric form for the oscillator
Hamiltonian~\eqref{e.Hs}, i.e., $a^2=b^2=\omega_0$; in this way both
$\gamma_{q}$ and $\gamma_{p}$ are dimensionless. Taking $\omega_0$ as
the frequency unit leads to a natural dimensionless formulation, with
the reduced temperature $t=1/(\beta\omega_0)$, the Matsubara
frequency $\nu_n=2\pi{tn}\omega_0$, and so on; all this is tantamount
to setting $\omega_0=1$, and in such units the classical part of the
dimensionless fluctuations is equal to the temperature,
$\big\langle{\hat{q}^2}\big\rangle_{\rm{cl}}\,{=}\,t$.

Therefore, the mean-square fluctuation we have to study is
\begin{equation}
 \big\langle{\hat q^2}\big\rangle_{\rm{pq}}
 \equiv \alpha(t,\gamma_{p},\gamma_{q},\tau_{_{\rm{D}}})~,
\end{equation}
which has the dimensionless explicit expression
\begin{equation}
 \alpha =
 2t \sum_{n=1}^\infty
 \frac{1+k_{pn}}{(2\pi{t}n)^2+(1+k_{pn})(1+k_{qn})}~,
\label{e.2b.q2}
\end{equation}
with
\begin{equation}
 k_{pn} =\gamma_{p}\,
 \frac{2\pi{t}n}{1+\tau_{_{\rm{D}}}2\pi{t}n}~,
~~~~
 k_{qn} =\gamma_{q}\,
 \frac{2\pi{t}n}{1+\tau_{_{\rm{D}}}2\pi{t}n}~.
\label{e.kpkqD}
\end{equation}
Note that a nonzero memory time $\tau_{_{\rm{D}}}$ is required to
make the series~\eqref{e.2b.q2} convergent.

\subsubsection{Zero temperature}
\label{sss.zeroT}

Taking the limit $t\to{0}$, one sets $x=2\pi{t}n$ and $dx=2\pi{t}$ to
write Eq.~\eqref{e.2b.q2} as an integral,
\begin{equation}
 \alpha(0,\gamma_{p},\gamma_{q},\tau_{_{\rm{D}}}) =
 \int_0^\infty \frac{dx}{\pi}\, \frac{1+\gamma_{p}\,g(x)}
 {\,x^2+[1{+}\gamma_{p}\,g(x)][1{+}\gamma_{q}\,g(x)]}~,
\label{e.2b.q2t0}
\end{equation}
where
\begin{equation}
 g(x)= \frac{x}{1+\tau_{_{\rm{D}}}x}~.
\label{e.gxDrude}
\end{equation}
From Eq.~\eqref{e.2b.q2t0} one can see that the result of the
isolated-system agrees with Eq.~\eqref{e.2b.p2q2pq0} in dimensionless
form,
\begin{equation}
 \alpha(0,0,0,\tau_{_{\rm{D}}})=\frac1\pi\int \frac{dx}{x^2+1}=\frac12~,
\end{equation}
and also recover a known analytic result~\cite{Weiss2008} in the case
of Ohmic standard dissipation,
\begin{eqnarray}
 \alpha(0,0,\gamma,0)
 =\frac2{\pi\sqrt{4{-}\gamma^2}}
 \tan^{-1}\frac{\sqrt{4{-}\gamma^2}}{\gamma}
 ~~(\gamma\,{<}\,2)~~&&
\notag \\
 =\frac2{\pi\sqrt{\gamma^2{-}4}}
 \tanh^{-1}\frac{\sqrt{\gamma^2{-}4}}{\gamma}
 ~~(\gamma\,{>}\,2).&&
\end{eqnarray}

\begin{figure}[t]
\includegraphics[width=85mm,angle=0]{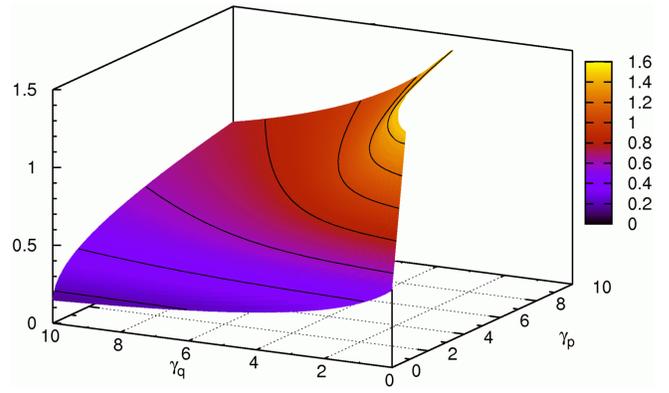}
\caption{Mean-square fluctuation of the coordinate at
$t\,{=}\,0$ as a function of both $\gamma_q$ and $\gamma_p$, for
$\tau_{_{\rm{D}}}\,{=}\,0.01$.}
\label{f.01}
\end{figure}

\begin{figure}[t]
\includegraphics[height=85mm,angle=90]{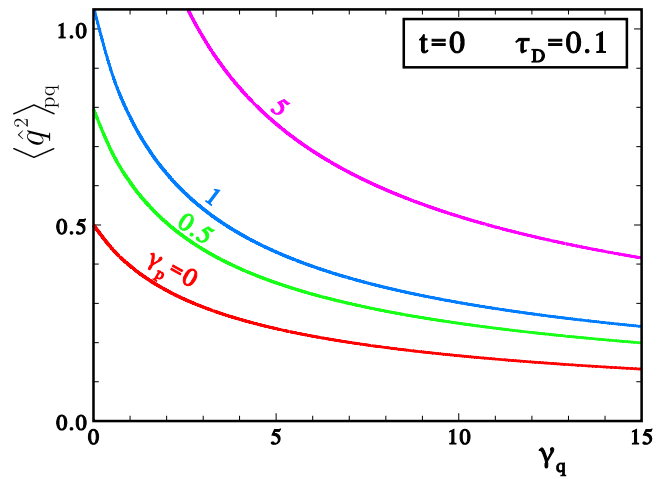}
\caption{Mean-square fluctuation of the coordinate at
$t\,{=}\,0$ as a function of $\gamma_q$ for fixed values of
$\gamma_p$. As in the case of standard dissipation,
$\gamma_p\,{=}\,0$, the effect of increasing the intensity of the
coordinate coupling is a decrease of $\langle{\hat{q}^2}\rangle$.}
\label{f.02}
\end{figure}

\begin{figure}[t]
\includegraphics[height=85mm,angle=90]{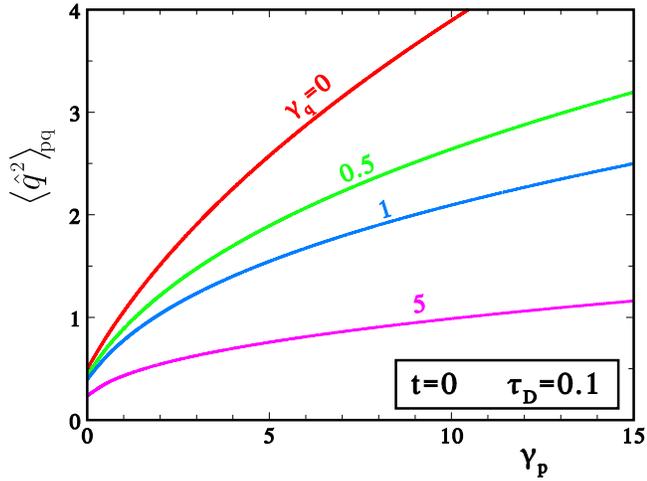}
\caption{Mean-square fluctuation of the coordinate at
$t\,{=}\,0$ as a function of $\gamma_p$ for fixed values of
$\gamma_q$. As in the case of anomalous dissipation,
$\gamma_q\,{=}\,0$, the effect of increasing the intensity of the
momentum coupling is an increase of $\langle{\hat{q}^2}\rangle$.}
\label{f.03}
\end{figure}

In the general case Eq.~\eqref{e.2b.q2t0} has to be evaluated
numerically. This is what we have done and Fig.~\ref{f.01} shows the
three-dimensional resulting plot for
$\alpha(0,\gamma_{p},\gamma_{q},\tau_{_{\rm{D}}}{=}0.01)$; its shape
gives an overall idea of the combined effect of two baths. However,
in order to highlight some important features which are barely
apparent, it is necessary to analyze its cross-sections.

In Fig.~\ref{f.02} we first consider how $\alpha$ varies while
increasing the coordinate coupling $\gamma_q$: also for nonzero
$\gamma_p$ the effect is a decrease of $\alpha$, as if the $q$-bath
were a position measuring device, causing localization of the
particle. From the same figure, and more clearly from
Fig.~\ref{f.03}, the complementary interpretation can be given to the
action of the $p$-bath, generalizing the observations made in the
case of \emph{anomalous} dissipation~\cite{CFTV2001,AnkerholdP2007};
note, however, that as soon as $\gamma_q$ is switched on the
steepness of the rise of $\alpha$ is considerably suppressed.

Given the opposite effect of the two baths, it is natural to ask what
the result will be when both have \emph{equal} strength,
$\gamma_p=\gamma_q$: this is a nontrivial issue, especially from the
point of view of the application to real problems where the canonical
variables play symmetric roles, as mentioned in the Introduction.

\begin{figure}[t]
\includegraphics[height=85mm,angle=90]{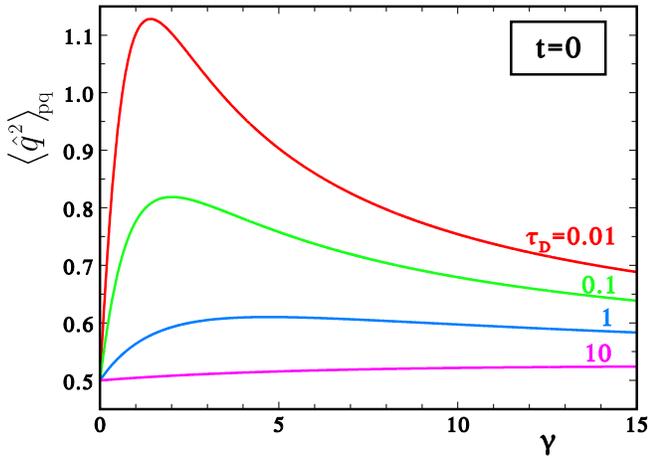}
\caption{Mean-square fluctuation of the coordinate at
$t\,{=}\,0$ as a function of $\gamma\,{=}\,\gamma_q\,{=}\,\gamma_p$,
i.e., for equally acting environments, for selected values of the
Drude time $\tau_{_{\rm{D}}}$. It appears that between the competing
effects of the two baths, the prevailing one is that of increasing
$\langle{\hat{q}^2}\rangle$; however, as the coupling intensity
increases a maximum is reached and followed by a slow decrease
towards the isolated-oscillator value.}
\label{f.04}
\end{figure}

\begin{figure}[t]
\includegraphics[width=85mm,angle=0]{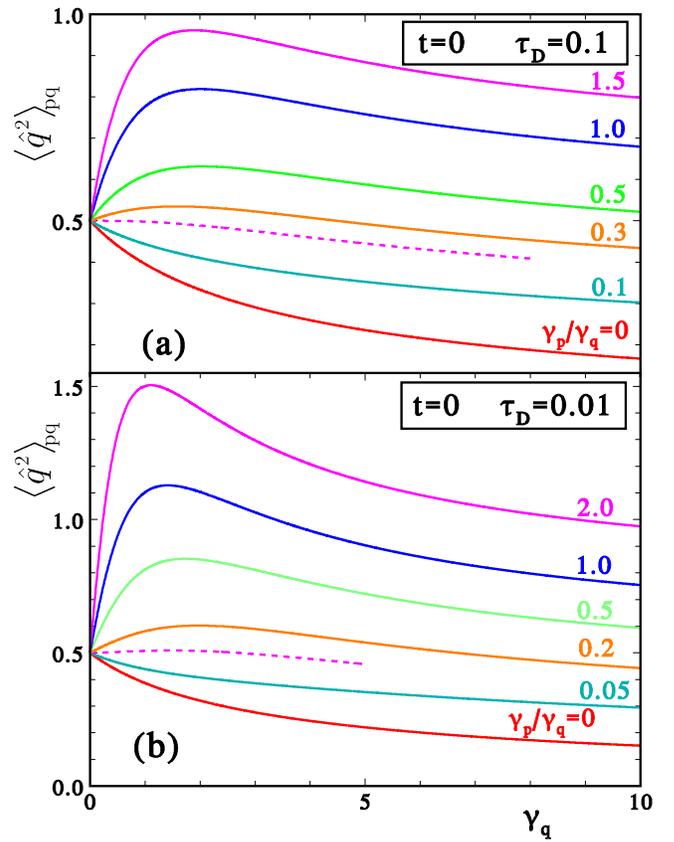}
\caption{Mean-square fluctuation of the coordinate for
$\tau_{_{\rm{D}}}\,{=}\,0.1$ (a) and $\tau_{_{\rm{D}}}\,{=}\,0.01$
(b) at $t\,{=}\,0$ along the line $\gamma_p\,{=}\,c\,\gamma_q$ as a
function of $\gamma_q$, for different values of $c$. The dashed line,
for $c{=}\,0.2214$ in (a) and $c{=}\,0.1193$ in (b), separates the
decreasing from the nonmonotonic behaviors.}
\label{f.05}
\end{figure}

\begin{figure}[t]
\includegraphics[height=85mm,angle=90]{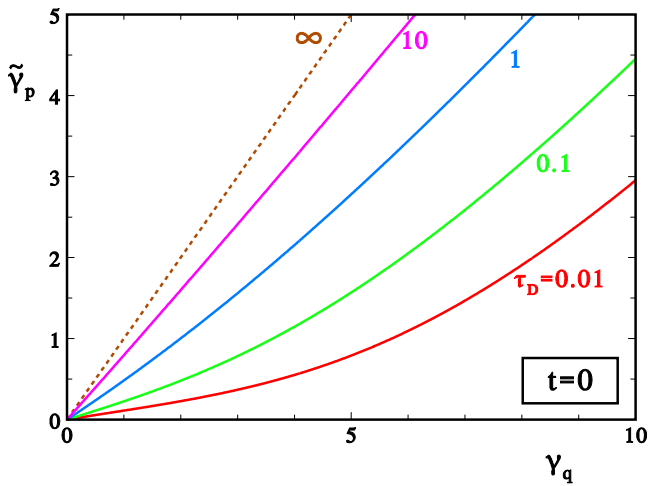}
\caption{The function $\tilde\gamma_p(\gamma_q)$
corresponding to the cancelation of the competing effects of the two
baths, so that the mean-square fluctuation of the coordinate keeps
the isolated system's value. The curves refer to $t\,{=}\,0$, so
$\langle{\hat{q}^2}\rangle=1/2$, and to selected values of the Drude
time $\tau_{_{\rm{D}}}$.}
\label{f.06}
\end{figure}

The answer is given by the curves reported in Fig.~\ref{f.04}:
$\alpha=\big\langle{\hat q^2}\big\rangle_{\rm{pq}}$ is larger
compared to the isolated oscillator, i.e., the quantum fluctuations
of both canonical variables are \emph{enhanced} by a symmetric
environmental coupling. However, an unexpected behavior shows up
while rising the intensity of the environmental coupling: after an
initial rise the mean-square fluctuation shows a maximum, whose
location and intensity depends upon the Drude memory time, and
eventually slowly decreases back towards the isolated oscillator
value $\alpha=1/2$.

It is obvious that, as the size of the zero-temperature fluctuations
is affected by the mechanism of environmental coupling, one could
formally modulate its intensity $\gamma$ to drive through a quantum
phase transition (QPT) a system that is close to it. At difference
with standard dissipative systems, e.g., resistively shunted
Josephson junction arrays, where a stronger dissipation leads towards
the `ordered' (superconducting) phase~\cite{Fisher1987}, in the
generalized case a higher $\gamma$ would most commonly increase
disorder. One can also speculate about the fact that the
nonmonotonicity displayed in Fig.~\ref{f.04} opens the possibility to
observe reentrant behavior around a QPT, namely by further rising
$\gamma$ beyond the critical value one could observe the restoration
of the isolated-system's ordered phase.

Having ascertained that the combined effect of the two baths is
opposite to that of the single $q$-bath, it is natural to look for
the evolution one observes by gradually switching the $p$-bath
coupling on. In Figs.~\ref{f.05} the behavior of $\alpha$ is followed
along straight lines in the plane $(\gamma_q,\gamma_p)$, i.e., for
fixed ratios $c=\gamma_p/\gamma_q$ between the intensities of the two
environmental couplings. These figures, which differ for the choice
of the Drude time, are essentially similar and display the competing
interplay between the intensities of the environmental couplings for
$q$ and $p$: if the latter is small enough, the quenching effect of
the former prevails. Increasing $\gamma_p$ for any fixed $\gamma_q$,
$\alpha$ must cross the isolated-system value $\alpha=1/2$ for some
$\tilde\gamma_p$: along the line $\tilde\gamma_p(\gamma_q)$ the
competing effects cancel each other and the mean-square fluctuation
of the coordinate is stationary; since $\alpha>1/2$ for
$\gamma_p=\gamma_q$ it follows that
$\tilde\gamma_p(\gamma_q)<\gamma_q$. From Eq.~\eqref{e.2b.q2t0} one
can analytically obtain the initial slope of these curves as a
function of the Drude memory time $\tau_{_{\rm{D}}}$,
\begin{equation}
 \tilde\gamma_p'(0)=
 \frac{2-\pi\tau_{_{\rm{D}}}-4\tau_{_{\rm{D}}}^2\ln\tau_{_{\rm{D}}}
                        +2\tau_{_{\rm{D}}}^2+\pi\tau_{_{\rm{D}}}^3}
      {-4\ln\tau_{_{\rm{D}}}-2+3\pi\tau_{_{\rm{D}}}
                        -2\tau_{_{\rm{D}}}^2+\pi\tau_{_{\rm{D}}}^3}~,
\end{equation}
which turns out to vanish in the Ohmic limit and tends to 1 for large
$\tau_{_{\rm{D}}}$. This is in agreement with the numerical
calculation of $\tilde\gamma_p(\gamma_q)$ reported in
Fig.~\ref{f.06}.

Fig.~\ref{f.05} also suggests that, keeping the ratio
$c={\gamma_p}/{\gamma_q}$ fixed, for large coupling intensity
$\langle{\hat{q}^2}\rangle$ tends to a finite value; one can indeed
find that
\begin{equation}
 \langle{\hat{q}^2}\rangle \to \frac12~\sqrt{c}
  = \frac12~\sqrt{\frac{\gamma_p}{\gamma_q}}~.
\end{equation}
Substantially, it turns out that, after a possible initial increase
of the spread of one of the canonical variables when the interaction
is switched on, a very strong environmental coupling eventually tends
to localize again both $\hat{p}$ and $\hat{q}$, as the above limit
also entails that
\begin{equation}
 \langle{\hat{p}^2}\rangle
 \to \frac12~\sqrt{\frac{\gamma_q}{\gamma_p}}~;
\end{equation}
the shrinking of the fluctuations is visibly more effective for the
variable with larger bath-coupling strength, but the product of the
fluctuations obeys the constraint of the uncertainty principle,
\begin{equation}
 \langle{\hat{q}^2}\rangle\langle{\hat{p}^2}\rangle \to \frac14~.
\end{equation}
In other words, a very large environmental coupling leads to a
situation where the uncertainty is {\it squeezed} along one axis in
the $p$-$q$ plane with respect to the symmetric spread in the free
system.

\begin{figure}[t]
\includegraphics[height=85mm,angle=90]{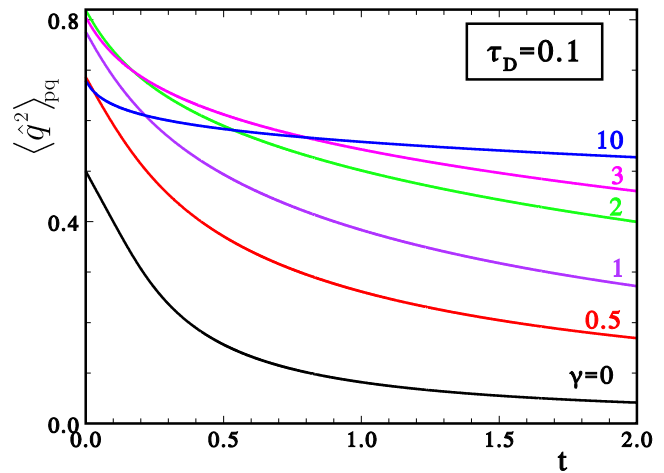}
\caption{Pure-quantum mean-square fluctuation of the
coordinate for $\tau_{_{\rm{D}}}\,{=}\,0.1$ and different values of
$\gamma\,{=}\,\gamma_q\,{=}\,\gamma_p$ (i.e., for equally acting
environments) as a function of  $t$.}
\label{f.07}
\end{figure}

\begin{figure}[t]
\includegraphics[height=85mm,angle=90]{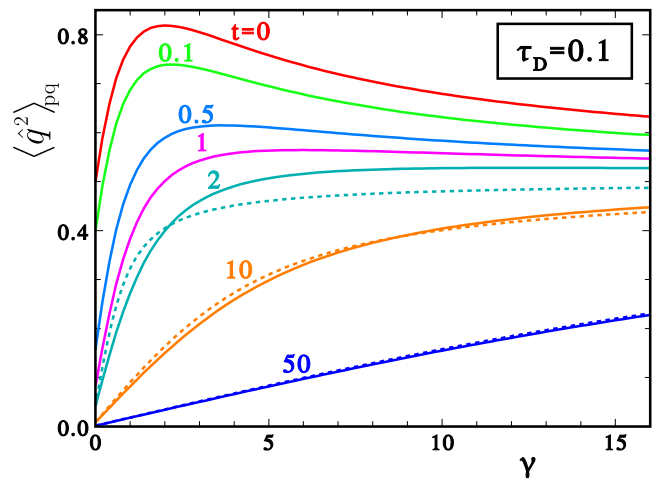}
\caption{Pure-quantum mean-square fluctuation of the
coordinate at different temperatures as a function of
$\gamma\,{=}\,\gamma_q\,{=}\,\gamma_p$, for $\tau_{_{\rm{D}}}=0.1$.
The dashed curves for $t\,{\ge}\,2$ represent the large-$t$ analytic
approximation~\eqref{e.alphatlarge} valid for
$t\gg1/(2\pi\tau_{_{\rm{D}}})\simeq1.6$.}
\label{f.08}
\end{figure}

\subsubsection{Finite temperature}
\label{sss.finiteT}

As Eq.~\eqref{e.2b.q2} shows, the effect of environmental coupling
only affects the purely quantum part of the fluctuations, which
decreases with increasing temperature: so, one expects a softening of
the effects observed at zero-$T$ when $T$ rises. However, the fate of
the reentrant fluctuation amplitude displayed in Figs.~\ref{f.04}
and~\ref{f.05} is not obvious. To unveil such issue we plot in
Fig.~\ref{f.07} the behavior of
$\alpha=\langle{\hat{q}^2}\rangle_{\rm{pq}}$ vs temperature for the
situation of equally acting environments. It appears that a stronger
environmental coupling weakens the decrease of
$\langle{\hat{q}^2}\rangle_{\rm{pq}}$ with temperature, and that for
$t\,{\gtrsim}\,1$ a monotonic behavior is apparently restored. This
fact is more evident by looking at the curves for increasing
temperatures reported in Fig.~\ref{f.08}.

When the temperature is larger than the Drude cutoff frequency,
$t\gg1/(2\pi\tau_{_{\rm{D}}})$, the expressions~\eqref{e.kpkqD} for
the kernels can be well approximated by their asymptotic values
$k_{qn}\simeq\gamma_{q}/\tau_{_{\rm{D}}}$ and
$k_{pn}\simeq\gamma_{p}/\tau_{_{\rm{D}}}$, so their dependence on $n$
disappears and the summation~\eqref{e.2b.q2} becomes analytic,
\begin{equation}
 \alpha\simeq
 \frac{c_p}{2c_q}\Big(\coth\frac{c_pc_q}{2t}-\frac{2t}{c_pc_q}\Big)
\label{e.alphatlarge}
\end{equation}
with $c_{p,q}^2=1+\gamma_{p,q}/\tau_{_{\rm{D}}}$: this approximation
is shown as dashed lines in Fig.~\ref{f.08}; for
$c_pc_q/{2t}\,{\gg}\,1$, i.e., under the combined condition
$\sqrt{\gamma_p\gamma_q}\gg 2\tau_{_{\rm{D}}}t \gg 1/\pi$, the value
of $\alpha$ tends to the constant ${c_p}/{2c_q}$. As the validity of
this approximation entails the disappearance of the maximum, the
nonmonotonic behavior is expected for
$t\,{\lesssim}\,1/(2\pi\tau_{_{\rm{D}}})$.

\section{Coordinate and momentum coupled with a single environment}
\label{s.1b}

It can also happen that there is one single environment coupled with
both canonical variables: the most general influence Hamiltonian can
be found in Ref.~\cite{KohlerS2006}. However, there are two main
scenarios of physical relevance: the first occurs when both $\hat{q}$
and $\hat{p}$ are coupled with the same bath variables (say, the
coordinates), while in the second the $q_\alpha$'s are coupled with
$\hat{q}$ and the $p_\alpha$'s with $\hat{p}$. The first scenario has
a realization in the coupling of spins with lattice vibrations due to
their modulation of the exchange integral; in the second case a
natural model could be a spin-orbit interaction where the orbital
angular momenta are thermalized by lattice vibrations.

\subsection{System coupled with bath coordinates}
\label{ss.1bq}

Starting from the coupling Hamiltonian,
\begin{equation}
 \hat{\cal{H}}_{\rm{I}} = \frac12\sum_\alpha\big[
 a_\alpha^2\hat p_\alpha^2
 + b_\alpha^2(\hat q_\alpha-c_\alpha\hat q-d_\alpha\hat p)^2\big]~,
\label{e.1bq.HI}
\end{equation}
we get the influence action:
\begin{equation}
 {\cal{S}}_{\rm{I}}[p,q] = -\frac{\beta}{2}\,\sum_n
 \big[k_{qn}~q_nq_{-n}+ k_{pn}~p_np_{-n}
 + 2\kappa_np_nq_{-n}\big]~,
\end{equation}
with the kernels
\begin{equation}
 k_{qn}=\sum_{\alpha}
 \frac{b_\alpha^2c_\alpha^2\nu_n^2}{\nu_n^2+\omega_\alpha^2}
~,~~~~
 k_{pn}=\sum_{\alpha}
 \frac{b_\alpha^2d_\alpha^2\nu_n^2}{\nu_n^2+\omega_\alpha^2}~,
\label{e.1bq.kqnkpn}
\end{equation}
\begin{equation}
 \kappa_n=\sum_{\alpha}
 \frac{b_\alpha^2c_\alpha d_\alpha\nu_n^2}{\nu_n^2+\omega_\alpha^2}~.
\label{e.1bq.kappan}
\end{equation}
It is immediate to check that the last kernel vanishes if
$c_\alpha{d_\alpha}=0$ for any $\alpha$, yielding the previous
Eq.~\eqref{e.SIpq}: indeed, if the coefficients $c_\alpha$ and
$d_\alpha$ are not simultaneously nonzero, the bath can be split into
two independent baths and the two-bath case studied in the previous
sections is recovered.

In the case when all the coefficients $c_\alpha$ and $d_\alpha$ are
positive the spectral density~\cite{Weiss2008} $J_{pq}(\omega)$ for
$\kappa_n$, such that
\begin{equation}
 \kappa_n = 2\nu_n^2 \int_0^\infty d\omega
\frac{J_{pq}(\omega)}{\omega^2+\nu_n^2}~,
\label{e.kappaJpq}
\end{equation}
can be expressed in terms of those for the first two kernels,
\begin{eqnarray}
 J_q(\omega) &\equiv& \sum_\alpha
 b_\alpha^2c^2_\alpha\omega_\alpha
 \,\delta(\omega^2{-}\omega_\alpha^2)~,
\notag\\
 J_p(\omega) &\equiv& \sum_\alpha
 b_\alpha^2d^2_\alpha\omega_\alpha
 \,\delta(\omega^2{-}\omega_\alpha^2)~,
\label{e.JqJp}
\end{eqnarray}
as their geometric average as their geometric
average~\cite{KohlerS2006}
\begin{equation}
 J_{pq}(\omega)=\sqrt{J_q(\omega)\,J_p(\omega)}~.
\label{e.Jpq}
\end{equation}
The baths are independent if the supports of $J_q(\omega)$ and
$J_p(\omega)$ have no intersection, i.e., if $J_{pq}(\omega)$
vanishes for any $\omega$.~\cite{footnote}

Eventually, using the result of Appendix~\ref{a.Gintegral}, the
expressions for the fluctuations in the case of the bath
coupling~\eqref{e.1bq.HI} are found,
\begin{equation}
 {\cal{Z}} = \frac1{\beta\omega_0}\prod_{n=1}^\infty
  \frac{\nu_n^2}{\nu_n^2+(a^2{+}k_{pn})(b^2{+}k_{qn})-\kappa_n^2},
\end{equation}

\begin{eqnarray}
 &&\big\langle{\hat p^2}\big\rangle_{\rm{pq}}
  = \frac2\beta\sum_{n=1}^\infty \frac{b^2{+}k_{qn}}
 {\nu_n^2+(a^2{+}k_{pn})(b^2{+}k_{qn})-\kappa_n^2}~,
\\
 &&\big\langle{\hat q^2}\big\rangle_{\rm{pq}}
  = \frac2\beta\sum_{n=1}^\infty \frac{a^2{+}k_{pn}}
 {\nu_n^2+(a^2{+}k_{pn})(b^2{+}k_{qn})-\kappa_n^2},
\notag\\
 &&\big\langle{\hat p\hat q+\hat q\hat p}\big\rangle_{\rm{pq}}
  = -\frac{4}\beta\sum_{n=1}^\infty  \frac{\kappa_n}
 {\nu_n^2+(a^2{+}k_{pn})(b^2{+}k_{qn})-\kappa_n^2}~.
\notag
\end{eqnarray}
As Eqs.~\eqref{e.1bq.kqnkpn} and~\eqref{e.1bq.kappan} imply the
inequality $k_{qn}k_{pn}\ge\kappa_n^2$, it follows that the
denominators are always positive. Note that if the bath is coupled to
the variable $\hat{p}\pm\hat{q}$ (i.e., $c_\alpha={\pm}d_\alpha$),
then $\pm\kappa_n=k_{pn}=k_{qn}>0$: from the above variances it
follows then that $\hat{p}\pm\hat{q}$ becomes smaller (larger) when
the environmental coupling is switched on,
\begin{equation}
 \big\langle{(\hat{p}\pm\hat{q})^2}\big\rangle_{\rm{pq}}
 = \frac2\beta\sum_{n=1}^\infty
 \frac{a^2+b^2{+}(k_{pn}+k_{qn}\mp2\kappa_n)}
 {\nu_n^2+(a^2{+}k_{pn})(b^2{+}k_{qn})-\kappa_n^2}~,
\label{e.1bq.ppmq2}
\end{equation}
which corresponds to a squeezed Gaussian distribution along a
diagonal in the $p$-$q$ plane, in agreement with the Heisenberg
uncertainty relation. This observation also confirms the overall
picture that the environment acts as a measuring device for whatever
variable it is coupled with.

Let us consider now a phenomenological guess fo the first two
kernels~\eqref{e.1bq.kqnkpn}, while the third one, $\kappa_n$,
follows from Eqs.~\eqref{e.kappaJpq}--\eqref{e.Jpq}. Taking Drude
kernels with equal memory times $\tau_{_{\rm{D}}}$ as in
Eq.~\eqref{e.kpkqDrude}, corresponding to the spectral densities
\begin{equation}
 J_{p,q}(\omega)=\frac{\gamma_{p,q}}\pi\,
                 \frac1{1+\tau_{_{\rm{D}}}^2\omega^2}~,
\label{e.JDrude}
\end{equation}
one simply finds
\begin{equation}
 \kappa_n =
 \frac{\sqrt{\gamma_q\gamma_p}~|\nu_n|}
  {1+\tau_{_{\rm{D}}}\,|\nu_n|}~;
\label{e.1bq.kappanDrude}
\end{equation}
in this case one exactly has $\kappa_n^2=k_{pn}k_{qn}$, and in the
dimensionless formulation introduced in Section~\ref{ss.2b.effects}
the coordinate pure-quantum fluctuation is given by
\begin{equation}
 \alpha =
 2t \sum_{n=1}^\infty
 \frac{1+k_{pn}}{(2\pi{t}n)^2+1+k_{pn}+k_{qn}}~,
\label{e.1bq.q2}
\end{equation}
with the kernels given in Eq.~\eqref{e.kpkqD}. For $t=0$ it becomes
the integral
\begin{equation}
 \alpha(0,\gamma_{p},\gamma_{q},\tau_{_{\rm{D}}}) =
 \int_0^\infty \frac{dx}{\pi}\, \frac{1+\gamma_{p}\,g(x)}
 {\,x^2+1{+}(\gamma_{p}{+}\gamma_{q})\,g(x)}~,
\label{e.1bq.q2t0}
\end{equation}
with $g(x)$ as in Eq.~\eqref{e.gxDrude}.

\begin{figure}[t]
\includegraphics[height=85mm,angle=90]{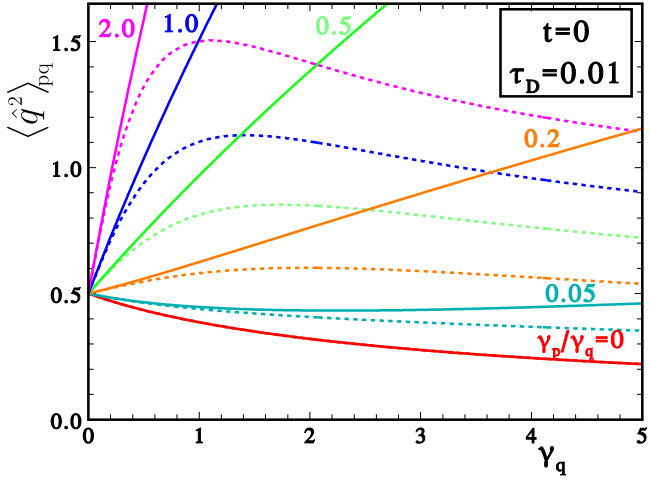}
\caption{System coupled with the bath coordinates.
Mean-square fluctuation of the coordinate, Eq.~\eqref{e.1bq.q2}, at
$t\,{=}\,0$ along the line $\gamma_p\,{=}\,c\,\gamma_q$ as a function
of $\gamma_q$, for different values of $c$. The dashed lines
correspond to the curves for the two-bath case of
Fig.~\ref{f.05}~(b).}
\label{f.09}
\end{figure}

In Fig.~\ref{f.09} it appears indeed that the mean-square fluctuation
of the coordinate is enhanced by the coupling mechanism considered in
this case, a result of the `diagonal squeezing' discussed after
Eq.~\eqref{e.1bq.ppmq2}. Of course, different distributions of the
coefficients $c_\alpha$ and $d_\alpha$ can yield a less dramatic
increase with the environmental coupling strength.

\subsection{System coupled with bath coordinates and momenta}
\label{ss.1b.pq-coupling}

In this case the most general interaction Hamiltonian would take the
form
\begin{equation}
 \hat{\cal{H}}_{\rm{I}} = \frac12\sum_\alpha\big[
 a_\alpha^2(\hat p_\alpha-c_\alpha\hat p)^2
 + b_\alpha^2(\hat q_\alpha-c_\alpha\hat q)^2\big]~;
\label{e.1bpq.HI}
\end{equation}
note that in this case the coefficients $c_\alpha$ cannot be absorbed
by means of linear canonical transformations of the bath variables.

The calculation of the influence action leads to an expression
similar to Eq.~\eqref{e.SIpq}, but for the appearance of an
antisymmetric correlation kernel $\nu_n\tilde\kappa_n$,
\begin{equation}
 {\cal{S}}_{\rm{I}}[p,q]
 = -\frac{\beta}{2}\sum_n \big(k_{pn}p_np_{-n} + k_{qn}q_nq_{-n}
 +2\nu_n\tilde\kappa_np_nq_{-n}\big),~
\label{e.1bpq.SIpq}
\end{equation}
with the Matsubara components
\begin{equation}
 k_{qn}=\sum_{\alpha}
 \frac{c_\alpha^2b_\alpha^2\nu_n^2}{\nu_n^2+\omega_\alpha^2}
~,~~~~
 k_{pn}=\sum_{\alpha}
 \frac{c_\alpha^2a_\alpha^2\nu_n^2}{\nu_n^2+\omega_\alpha^2}~,
\label{e.1bpq.kqnkpn}
\end{equation}
\begin{equation}
 \tilde\kappa_n=\sum_{\alpha}
 \frac{c_\alpha^2\omega_\alpha^2}{\nu_n^2+\omega_\alpha^2}~.
\label{e.1bpq.kappan}
\end{equation}
Using the result of Appendix~\ref{a.Gintegral} one has the
expressions for the partition function and the mean-square
fluctuations:
\begin{equation}
 {\cal{Z}} = \frac1{\beta\omega_0}\prod_{n=1}^\infty \frac{\nu_n^2}
 {\nu_n^2(1{+}\tilde\kappa_n)^2+(a^2{+}k_{pn})(b^2{+}k_{qn})}~,
\label{e.1bpq.Z}
\end{equation}
\begin{eqnarray}
 \big\langle{\hat p^2}\big\rangle_{\rm{pq}}
  &=& \frac2\beta\sum_{n=1}^\infty \frac{b^2+k_{qn}}
  {\nu_n^2(1{+}\tilde\kappa_n)^2+(a^2{+}k_{pn})(b^2{+}k_{qn})}~,
\notag\\
 \big\langle{\hat q^2}\big\rangle_{\rm{pq}}
  &=& \frac2\beta\sum_{n=1}^\infty \frac{a^2+k_{pn}}
  {\nu_n^2(1{+}\tilde\kappa_n)^2{+}(a^2{+}k_{pn})(b^2{+}k_{qn})}\,,~
\label{e.1bpq.q2p2}
\end{eqnarray}
while $\langle{\hat{p}\hat{q}+\hat{q}\hat{p}}\rangle=0$. Therefore,
the kernel $\tilde\kappa_n$ has the effect to lower the fluctuations
of both canonical variables with respect to the case of two
independent baths. It is reasonable to use a phenomenological guess
for $k_{qn}$ and $k_{pn}$, while $\tilde\kappa_n$ is connected with
them: in terms of the spectral densities it can be represented as
\begin{equation}
 \tilde\kappa_n = 2 \int_0^\infty
 \frac{\omega \sqrt{J_q(\omega)\,J_p(\omega)}}{\omega^2+\nu_n^2}~
  d\omega~.
\end{equation}
Taking Drude kernels with equal memory times $\tau_{_{\rm{D}}}$ as in
Eq.~\eqref{e.kpkqDrude}, corresponding to the spectral
densities~\eqref{e.JDrude}, one obtains
\begin{equation}
 \tilde\kappa_n =  \sqrt{\gamma_q\gamma_p}~ f(\tau_{_{\rm{D}}}\nu_n)~,
~~~~~~
  f(x) \equiv \frac1\pi~\frac{\ln x^2}{x^2-1}~.
\label{e.1bpq.kappanDrude}
\end{equation}
Note that $\tilde\kappa_n$ is positive and that
$\nu_n\tilde\kappa_n\sim(\ln{n})/n\to0$ for $n\to\infty$. In the
dimensionless formulation of Section~\ref{ss.2b.effects}, with the
kernels~\eqref{e.kpkqD}, the pure-quantum fluctuation of the
coordinate reads
\begin{equation}
 \alpha = 2t \sum_{n=1}^\infty \frac{1+k_{pn}}
 {(2\pi{t}n)^2(1{+}\tilde\kappa_n)^2+(1{+}k_{pn})(1{+}k_{qn})}~,
\label{e.1bpq.q2}
\end{equation}
with
$\tilde\kappa_n=\sqrt{\gamma_q\gamma_p}~f(2\pi{t}n\tau_{_{\rm{D}}})$.
At zero temperature this becomes the integral
\begin{eqnarray}
 && \alpha(0,\gamma_{p},\gamma_{q},\tau_{_{\rm{D}}}) =
\label{e.1bpq.q2t0}
\\ \notag
 ~~&& =\int_0^\infty \frac{dx}{\pi}\, \frac{1+\gamma_{p}\,g(x)}
 {x^2[1{+}\tilde\kappa(x)]^2
  +[1{+}\gamma_{p}\,g(x)][1{+}\gamma_{q}\,g(x)]}~,
\end{eqnarray}
with $\tilde\kappa(x)=\sqrt{\gamma_q\gamma_p}\,f(\tau_{_{\rm{D}}}x)$
and $g(x)$ as in Eq.~\eqref{e.gxDrude}

\begin{figure}[t]
\includegraphics[height=85mm,angle=90]{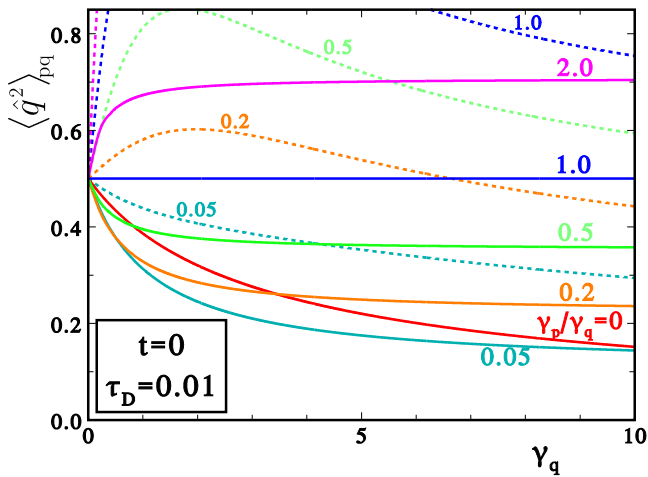}
\caption{System coupled with the bath coordinates and
momenta. Mean-square fluctuation of the coordinate,
Eq.~\eqref{e.1bpq.q2}, at $t\,{=}\,0$ along the line
$\gamma_p\,{=}\,c\,\gamma_q$ as a function of $\gamma_q$, for
different values of $c$. The dashed lines correspond to the curves
for the two-bath case of Fig.~\ref{f.05}~(b).}
\label{f.10}
\end{figure}

From Fig.~\ref{f.10} one can see that an environmental coupling of
the kind of Eq.~\eqref{e.1bpq.HI} produces much smaller effects than
the previously considered ones. Note that the effect of increasing
the bath-momentum coupling $\gamma_p$ even initially shrinks the
coordinate fluctuations, with a further slow rise. Moreover, it is
particularly surprising to see that the existence of environmental
coupling is irrelevant in the symmetric case; this means that the
integral~\eqref{e.1bpq.q2t0} sticks to the isolated system's
value~$1/2$, whatever the values of $\gamma_p\,{=}\,\gamma_q$ and
$\tau_{_{\rm{D}}}$: this result is far from being apparent from
Eq.~\eqref{e.1bpq.q2t0}.

\section{Conclusions}
\label{s.concl}

Motivated by quest for a suitable description of spin-lattice
coupling, with the lattice regarded as a harmonic bath of thermalized
degrees of freedom, we considered a generalized
\emph{system-plus-reservoir} (SPR) model where both coordinates and
momenta are subject to the interaction with the environment. While it
was known that coupling a bath to a canonical variable quenches its
fluctuations, attaching independent baths to both coordinate and
momentum generally enhances their fluctuations. In the case of a spin
system, such larger quantum fluctuations could open the possibility
of reaching the quantum phase transition of the spin-$\frac12$
two-dimensional Heisenberg antiferromagnet~\cite{ChakravartyHN1989},
i.e., spin-lattice coupling could yield a vanishing zero-$T$
sublattice magnetization with a power-law divergent correlation
length. This possibility is to be further analyzed by a
quantitatively accurate approach. Moreover and surprisingly, it turns
out that the effect is nonmonotonic with the coupling strength,
opening the possibility of a reentrant behavior in such a quantum
critical system driven by its environmental coupling strength.

Other kinds of coupling involving both canonical variables and a
single bath have been considered and briefly discussed. If the
coupling terms do not simultaneously involve noncommuting bath
variables (e.g., only bath coordinates) the result is a modification
of the two-bath scenario with a further enhancement of the
fluctuations and disappearance of the above mentioned
nonmonotonicity. In the opposite case it clearly appears that the
further `interference' effect gives rise to generally smaller
fluctuations, i.e., the intensity of bath-momentum coupling is much
less effective in increasing the coordinate fluctuations, which
moreover show two exotic features: a counterintuitive initial
decrease when $\gamma_p$ is switched on and the intriguing stability
to the isolated system's value in the symmetric case
$\gamma_p\,{=}\,\gamma_q$; in other words, a symmetric environmental
interaction appears to be ineffective for what attains mean square
fluctuations.

\acknowledgements
We thank Prof. Valerio Tognetti for stimulating us to further
investigate these topics and for his useful suggestions.

\appendix

\section{The standard system-plus-reservoir model}
\label{a.standard}

A bath attached to the coordinate $\hat{q}$ of the system of
interest, e.g., the oscillator~\eqref{e.Hs}, is a collection of
harmonic oscillators described by the Hamiltonian
\begin{equation}
 \hat{\cal{H}}_{\rm{I}} =\frac12\sum_\alpha\big[
 a_\alpha^2\hat p_\alpha^2 + b_\alpha^2(\hat q_\alpha-\hat q)^2\big]~.
\label{e.Hb}
\end{equation}
Its characteristics are `microscopically' specified by the collection
of positive parameters $\{a_\alpha,b_\alpha\}$, the corresponding
frequencies being $\omega_\alpha\,{=}~a_\alpha\,b_\alpha$. At
variance with the common use~\cite{Weiss2008} we do not include here
a third set of parameters $c_\alpha$ in front of $\hat{q}$, because
they can be absorbed by means of a canonical transformation which
affects the bath variables only,
$\hat{q}_\alpha\,{\to}\,c_\alpha\,\hat{q}_\alpha$,
$\hat{p}_\alpha\,{\to}\,c_\alpha^{-1}\,\hat{p}_\alpha$.

For the quantum thermodynamics at the equilibrium temperature
$T=1/\beta$, the \emph{influence action} ${\cal{S}}_{\rm{I}}[q]$
corresponding to the Hamiltonian~\eqref{e.Hb} is obtained by
\emph{tracing out} the bath variables, i.e., by integrating over them
in the corresponding path integral. The influence action is bi-local
in imaginary time (a local contribution would arise from a
Hamiltonian operator for the system and couldn't describe
dissipation), but it is local (or diagonal) in Matsubara
space~\cite{Weiss2008}:
\begin{eqnarray}
 {\cal{S}}_{\rm{I}}[q]
  &=& -\frac12\int_0^\beta du\,du'~k(u{-}u')~q(u)q(u')~,
\notag\\
 &=& -\frac{\beta}{2}\,\sum_n k_n~q_nq_{-n}~,
\label{e.SIq}
\end{eqnarray}
with the components $q_n$ for each Matsubara frequency
$\nu_n\equiv2\pi\,n/\beta$ defined by
$q(u)\equiv\sum_nq_n\,e^{-i\nu_nu}$.
The Matsubara components of the \emph{kernel} $k(u)$ are related to
the bath parameters through
\begin{equation}
 k_n\equiv\int_0^\beta du~e^{i\nu_nu}k(u)
 =\nu_n^2 \sum_{\alpha} \frac{b_\alpha^2}{\nu_n^2+\omega_\alpha^2}~.
\end{equation}
Starting from the Hamiltonian
$\hat{\cal{H}}=\hat{\cal{H}}_{\rm{S}}+\hat{\cal{H}}_{\rm{I}}$, after
eliminating the bath variables from the equations of motion one finds
that the dynamics is ruled by a quantum Langevin
equation~\cite{FordLO1988}, where dissipation is described in terms
of a (retarded) \emph{memory function} $\gamma(t)$ whose Laplace
transform is expressed in terms of the microscopic bath parameters as
\begin{equation}
 \gamma(z)=\int_0^\infty dt~e^{-zt}\gamma(t)
  = z \sum_\alpha  \frac{b_\alpha^2}{z^2+\omega_\alpha^2}~;
\label{e.gamma}
\end{equation}
hence, it is apparent that the kernel is related to the memory
function,
\begin{equation}
 k_n  =|\nu_n| \,\gamma(z{=}|\nu_n|)~.
\label{e.kngamma}
\end{equation}
This equality is a valuable relation, as it connects the quantum
dissipative kernel with a phenomenological quantity that can be
experimentally accessed in a dissipative dynamical process.

The full system's action, i.e., the sum of the action
${\cal{S}}_{\rm{S}}$ of the isolated oscillator and the influence
action ${\cal{S}}_{\rm{I}}$, is
\begin{equation}
 {\cal{S}} = -\frac\beta2\sum_n \big[ 2\nu_n p_nq_{-n}
 {+}\,a^2p_np_{-n} {+}(b^2{+}k_n)q_nq_{-n} \big],
\end{equation}
and using the general result of Appendix~\ref{a.Gintegral} one finds
the known results~\cite{Weiss2008} for the partition function,
\begin{equation}
 {\cal{Z}}=\frac1{\beta\omega_0}\prod_{n=1}^\infty
 \frac{\nu_n^2}{\nu_n^2+a^2(b^2{+}k_n)}~,
\end{equation}
and for the mean-square fluctuations of momentum and coordinate,
\begin{eqnarray}
 \big\langle{\hat p^2}\big\rangle &=&
 \frac1\beta\sum_n\frac{b^2+k_n}{\nu_n^2+a^2(b^2{+}k_n)}~,
\notag \\
 \big\langle{\hat q^2}\big\rangle &=&
 \frac1\beta\sum_n\frac{a^2}{\nu_n^2+a^2(b^2{+}k_n)}~.
\label{e.2b.p2q2standard}
\end{eqnarray}
The above known results~\cite{Weiss2008} qualitatively mean that
standard dissipation decreases the quantum fluctuations of the
coordinate, while it increases those of the momentum, compared to the
limit of no dissipation ($k_n\to{0}$), where the usual quantum
expressions are recovered, i.e.,
${\cal{Z}}=[2\sinh(\beta\omega_0/2)]^{-1}$ and
$\langle{\hat{p}^2}\rangle/b^2
=\langle{\hat{q}^2}\rangle/a^2=(1/2\omega_0)\coth(\beta\omega_0/2)$.
Note that in the \emph{Ohmic} case, where $k_n\sim{n}$ for large $n$,
the fluctuation of $\hat{p}$ diverges, while that of $\hat{q}$
remains finite.

Of course, the symmetry of interchanging the canonical variables
tells us that when the bath is coupled with the momentum, i.e., if
\begin{equation}
 \hat{\cal{H}}_{\rm{I}} =\frac12\sum_\alpha\big[
 a_\alpha^2(\hat p_\alpha-\hat p)^2 + b_\alpha^2\hat
 q_\alpha^2\big]~,
\end{equation}
the exact converse occurs~\cite{Leggett1984,CFTV2001,AnkerholdP2007}:
for the harmonic oscillator this is immediately clear from the fact
that the canonical transformation $\hat{p}\to-\hat{q}$ and
$\hat{q}\to\hat{p}$ maps the two coupling models onto each other.

\section{The general Gaussian integral}
\label{a.Gintegral}

Let us consider the following general quadratic action in Matsubara
space,
\begin{equation}
 {\cal{S}} = -\frac\beta2\sum_n
  ~\big[ a^2_n\,p_np_{-n} + b^2_n\, q_nq_{-n}
 + 2(\pi_n{+}\,\delta_n)\, p_nq_{-n} \big]~,
\label{e.a.Squadratic}
\end{equation}
where $a_n=a_{-n}$, $b_n=b_{-n}$, and $\pi_n=\pi_{-n}$ are positive,
while the antisymmetric part $\delta_n=-\delta_{-n}$ is such that
$\delta_n\sim\nu_n$, the Matsubara frequency, for large $n$. The
Gaussian average of any quantity $O\big(\{p_n,q_n\}\big)$ is
\begin{equation}
 \langle{O}\rangle = \frac J{\cal{Z}}
        \prod_n \int \frac{dp_ndq_n}{2\pi}
        ~O\big(\{p_n,q_n\}\big)~e^{{\cal{S}}}~,
\end{equation}
where $J=\prod_{n=1}^\infty(\beta\nu_n)^2$ is the Jacobian of the
Matsubara transformation, ${\cal{Z}}$ is the partition function,
obtained for $O=1$, and the integrals are defined as ordinary
integrals over the independent real and imaginary parts as
\begin{equation}
 \prod_n\int dp_n \equiv
 \int dp_0 \prod_{n=0}^\infty 2\int dp_n^{\rm{R}}\,dp_n^{\rm{I}}~,
\end{equation}
where $p_{\pm{n}}\,{\equiv}\,p_n^{\rm{R}}\,{\pm}\,i\,p_n^{\rm{I}}$,
and the analog for the coordinates.

The action can be easily worked out by means of simple
transformations that preserve the measure:

(\emph{i})~ the coefficients of momenta and coordinates are made
equal to $\omega_n=a_nb_n$ by
\begin{equation}
 p_n = \sqrt{\frac{b_n}{a_n}}~p_n^{_{(1)}}~,
~~~~
 q_n = \sqrt{\frac{a_n}{b_n}}~q_n^{_{(1)}}~,
\end{equation}

(\emph{ii})~ the terms that multiply $\pi_n$ are diagonalized to
$p_n^{_{(2)}}p_{-n}^{_{(2)}}-q_n^{_{(2)}}q_{-n}^{_{(2)}}$ by the
rotation
\begin{equation}
 \left(\begin{array}{c}p_n^{_{(1)}}\\q_n^{_{(1)}}\end{array}\right)
 = \frac1{\sqrt2}\left(\begin{array}{rr} 1 & -1
                                      \\ 1 &  1 \end{array}\right)
 \left(\begin{array}{c}p_n^{_{(2)}}\\q_n^{_{(2)}}\end{array}\right)~,
\end{equation}

(\emph{iii})~ so that a further balancing is in order,
\begin{equation}
 p_n^{_{(2)}}
 = \sqrt[4]{\frac{\omega_n-\pi_n}{\omega_n+\pi_n}}~\tilde p_n~,
~~~~
 q_n^{_{(2)}}
 = \sqrt[4]{\frac{\omega_n+\pi_n}{\omega_n-\pi_n}}~\tilde q_n~;
\end{equation}
eventually giving
\begin{eqnarray}
 {\cal{S}} &=& -\frac\beta2\sum_n ~\big[\Omega_n
 (\tilde p_n\tilde p_{-n}{+}\,\tilde q_n\tilde q_{-n})
  + 2\delta_n\, \tilde p_n\tilde q_{-n}\big]
\notag\\
 &=& -\frac\beta2\sum_n ~\Big(\Omega_n\, \tilde p'_n\tilde p'_{-n}
  + \frac{\delta_n^2+\Omega_n^2}{\Omega_n}~\tilde q_n\tilde q_{-n}
   \Big)
\end{eqnarray}
with $\Omega_n^2\equiv\omega_n^2-\pi_n^2=a_n^2b_n^2-\pi_n^2$ and
$\tilde{p}'_n=\tilde{p}_n-(\delta_n/\Omega_n)\,\tilde{q}_n$; note
that for $\Omega_n$ to be real one must require
\begin{equation}
 \pi_n<\omega_n~.
\end{equation}
It is now straightforward to obtain the variances
\begin{eqnarray}
 \langle{\tilde p_n\tilde p_{-n}}\rangle
 = \langle{\tilde q_n\tilde q_{-n}}\rangle
  &=& \frac1\beta\, \frac{\Omega_n}{\delta_n^2+\Omega_n^2}
\notag\\
 \langle{\tilde p_n\tilde q_{-n}}\rangle
  &=& \frac1\beta\, \frac{\delta_n}{\delta_n^2+\Omega_n^2}~.
\end{eqnarray}
Integrating over $\tilde{p}'_n$ and $\tilde{q}_n$ one finds the
partition function
\begin{equation}
 {\cal{Z}} = \frac1{\beta\Omega_0}\prod_{n=1}^\infty
  \frac{\nu_n^2}{\delta_n^2+\Omega_n^2}~.
\label{e.a.Z}
\end{equation}
It is easy to go backwards through the transformations (\emph{iii}),
(\emph{ii}), and (\emph{i}) to obtain the explicit result for the
original Matsubara variables of Eq.~\eqref{e.a.Squadratic},
\begin{eqnarray}
 \langle{p_np_{-n}}\rangle &=&
  \frac1\beta\, \frac{b_n^2}{\delta_n^2+a_n^2b_n^2-\pi_n^2}
\notag\\
 \langle{q_nq_{-n}}\rangle &=&
  \frac1\beta\, \frac{a_n^2}{\delta_n^2+a_n^2b_n^2-\pi_n^2}
\notag\\
 \langle{p_nq_{-n}}\rangle &=&
  \frac1\beta\,\frac{\delta_n-\pi_n}{\delta_n^2+a_n^2b_n^2-\pi_n^2}~.
\label{e.a.variances}
\end{eqnarray}


\begin{thebibliography}{00}

\bibitem{Weiss2008}
 U. Weiss, {\em Quantum Dissipative Systems}
 (3rd ed., World Scientific, Singapore, 2008).

\bibitem{Ullersma1966}
 P.~Ullersma, Physica (Amsterdam) {\bf 32}, 27 (1966);
  {\bf 32}, 56 (1966); {\bf 32}, 74 (1966); {\bf 32}, 90 (1966).

\bibitem{CaldeiraL1983}
 A.~O. Caldeira and A.~J. Leggett,
 Ann. of Phys. {\bf 149}, 374 (1983).

\bibitem{CRTV1997}
 A. Cuccoli, A. Rossi, V. Tognetti, and R. Vaia,
 Phys. Rev. E {\bf 55}, R4849 (1997).

\bibitem{CFTV1999}
 A. Cuccoli, A. Fubini, V. Tognetti, and R. Vaia,
 Phys. Rev. E {\bf 60}, 231 (1999).

\bibitem{FordLO1988}
 G.~W. Ford, J.~T. Lewis, and R.~F. O'Connell,
 Phys. Rev. A {\bf 37}, 4419 (1988).

\bibitem{HaseTU1993}
 M. Hase, I. Terasaki, and K. Uchinokura
 Phys. Rev. Lett. {\bf 70}, 3651 (1993).

\bibitem{BursillMH1999}
 R.~J. Bursill, R.~H. McKenzie, and C.~J. Hamer,
 Phys. Rev. Lett. {\bf 83}, 408 (1999).

\bibitem{CTVV1995xxz}
 A. Cuccoli, V. Tognetti, P. Verrucchi, and R. Vaia,
 Phys. Rev. B {\bf 51},  12840 (1995).

\bibitem{CRTVV2000}
 A. Cuccoli, T. Roscilde, V. Tognetti, P. Verrucchi, and R. Vaia,
 Phys. Rev. B {\bf 62}, 3771 (2000).

\bibitem{CRTVV2001}
 A.Cuccoli, T.Roscilde, V.Tognetti, R.Vaia, P.Verrucchi
 Eur. Phys. J. B {\bf 20}, 55 (2001).

\bibitem{CFTV2008}
 A. Cuccoli, A. Fubini, V. Tognetti, and R. Vaia,
 in \emph{Path Integrals - New Trends and Perspectives},
 edited by W. Janke and A. Pelster
 (World Scientific, Singapore, 2008), p. 500.

\bibitem{Leggett1984}
 A.~J. Leggett, Phys. Rev. B {\bf 30}, 1208 (1984).

\bibitem{CFTV2001}
 A. Cuccoli, A.~Fubini, V. Tognetti, and R. Vaia,
 Phys. Rev. E {\bf 64}, 066124 (2001).

\bibitem{AnkerholdP2007}
 J. Ankerhold and E. Pollak,
 Phys. Rev. E {\bf 75}, 041103 (2007).

\bibitem{KohlerS2005}
 H. Kohler and F.~Sols,
 Phys. Rev. B {\bf 72}, 180404(R) (2005).

\bibitem{NovaisCBAZ2005}
 E. Novais, A.~H. Castro Neto, L. Borda, I. Affleck, and G. Zar\'and,
 Phys. Rev. B {\bf 72}, 014417 (2005).

\bibitem{KohlerS2006}
 H. Kohler and F.~Sols,
 New J. of Phys. {\bf 8}, 149 (2006).

\bibitem{HaakeR1984}
 F. Haake, R. Reibold,,
 Phys. Rev. A {\bf 32}, 2462 (1985).

\bibitem{Talkner1986}
 P. Talkner,
 Ann. Phys. (N.Y.) {\bf 167}, 390 (1986).

\bibitem{Fisher1987}
 M.~P.~A. Fisher,
 Phys. Rev. B {\bf 36}, 1917 (1987).

\bibitem{footnote}
The equivalence between the independent-bath condition and
non-intersecting supports of $J_q(\omega)$ and $J_p(\omega)$ clearly
shows that the treatment made in Section~\ref{ss.2b.effects} cannot
be obtained as a particular case of what we are discussing here: the
continuum limit underlying Eq.~\eqref{e.2b.q2t0} implies indeed that
the supports of $J_q(\omega)$ and $J_p(\omega)$ do overlap.

\bibitem{ChakravartyHN1989}
 S. Chakravarty, B.~I. Halperin, and D.~R. Nelson,
 Phys. Rev. B {\bf 39},  2344 (1989).

\end{thebibliography}
\end{document}